\documentclass[twocolumn,twoside,9pt]{article}
\usepackage{epsfig,amsmath,amsfonts,amssymb,multirow,textcomp}
\usepackage[T1]{fontenc}

\usepackage{graphicx}
\newcommand{\fd}{\mbox{$.\!^{\rm d}$}}
\newcommand{\fm}{\mbox{$.\!^{\rm m}$}}
\makeatother
\begin{document}
\title{VARIABILITY OF THE SPIN PERIOD\\ 
OF THE WHITE DWARF\\ 
IN THE INTERMEDIATE POLAR V405 AUR:\\
LOW-MASS THIRD BODY OR PRECESSION ?}
\author{V. V. Breus$^{1}$, I. L. Andronov$^{1}$, P. Dubovsk\'y$^{2}$,\\
S. V. Kolesnikov$^{3}$, E. A. Zhuzhulina$^{4}$, T. Heged\"us$^{5}$,\\
 P. Beringer$^{5}$, K. Petr\'ik$^{6}$, J. W. Robertson$^{7}$, A.V. Ryabov$^{3}$, I. Kudzej$^{2}$, 
\fbox{N. M. Shakhovskoy}$^8$\\[2mm]
$^1$ Department "High and Applied Mathematics", Odessa National Maritime University\\ 
Odessa, Ukraine, {e-mail \em vitaly.breus@gmail.com, tt\_ari@ukr.net}\\
$^2$ Vihorlat Astronomical Observatory, Humenne, Slovakia\\
$^3$ Astronomical Observatory,  Odessa National University, Odessa, Ukraine\\
$^4$ Department of Astronomy, Kharkiv National University, Kharkiv, Ukraine\\
$^5$ Baja Astronomical Observatory, Baja, Hungary\\
$^6$ Astronomical Observatory, Hlohovec, Slovakia\\
$^7$ Arkansas Tech University, Russellville, USA\\
$^8$ Crimean Astrophysical Observatory, Nauchny, Ukraine\\
}
\date{\em Submitted to ``Journal of Physical Studies", on June 10, 2013, accepted, vol. 17}
\maketitle

\begin{abstract}
We present the results of photometric CCD observations of the magnetic cataclysmic variable V405 Aurigae (RX J0558.0+5353 = 1RXS J055800.7+535358) obtained using different instruments. We analysed variability of the spin period of the white dwarf in the V405 Aur (RX J0558.0+5353) system using our observations and previously published maxima timings. The spin period of the system in 2010-2012 is $P=545.4558163(94)$s.
As we have gaps in observational data, we present 2 hypotheses of the spin period variability of this system - a cubic ephemeris which may be interpreted by a precession of the magnetic white dwarf or a periodic change with a period of 6.2 years and semi-amplitude of $17.2\pm1.8$ sec. The periodic variations may be interpreted by a light-time effect caused by a low-mass star ($M_3\ge0.09M_\odot$). In this case, the system belongs to a rare class of cataclysmic variables with a third body.\\[1ex]
\end{abstract}

\section*{ Introduction}
\indent \indent The intermediate polar V405 Aur  was discovered as an optical counterpart of the ROSAT soft X-RAy source 1RXS J055800.7+535358 (RX J0558.0+5353) by Haberl et al. \cite{haberl94}. The soft X-Ray flux was changing with a period of 272.74s, which was supposed to be a spin period of the white dwarf. The presence of optical pulsations at a period of $272.785\pm0.003$s was reported by Ashoka et al. \cite{ashoka95}.

Later Allan et al. \cite{allan95} and Skillman \cite{skillman96} made independent announcements that the spin period of the white dwarf is twice that previously suggested (545.45s). The double value of the spin period was justified by detection of circular polarization with a period of $P =
0.006301\pm0.000055$d $(544.4\pm4.8)$s and semi-amplitude of
$1.80\pm0.16$ percent (Shakhovskoj and Kolesnikov \cite{shakh97}). Noting very similar photometric maxima (two during a polarimetric period), Shakhovskoj et al. \cite{shakh01} suggested a nearly-equatorial location of two accretion columns.

Evans and Hellier \cite{evans04} discussed the double-peaked spin pulse. The shape of the phase curve is wavelength-dependent, with a strong separation between the peaks at soft X-Ray range and a "saw-tooth" shape at hard X-Rays. This is naturally explained by an inequality of the columns.

Piirola et al. \cite{piirola08} analysed the maxima timings obtained in 1994-2007 and published second-order polynomial fit to the timings:
\begin{eqnarray}\label{form1}
T_{max}&=& HJD~2449681.46389(5)\nonumber\\
&& + 0.0063131474(4)E \\
&&+ 4(4)\cdot10^{-16}E^{2}\nonumber
\end{eqnarray}
The quadratic term is formally positive (corresponding to a period {\em increase).} However, this term deviates from zero by $1\sigma,$ so is not statistically significant. In brackets is a statistical error estimate in units of a last digit.

\section*{Observations}
\indent \indent We obtained photometric CCD observations using different telescopes: 1m VNT (filters V and R) and 28cm Pupava (unfiltered) reflectors in Vihorlat Astronomical Observatory, Humenn\'e, Slovakia, 35cm BAT and 50cm reflector in Baja Astronomical Observatory, Hungary (V and R filters) and 20cm MEADE LX-200 at the Observatory and Planetarium in Hlohovec, Slovakia (V and R filters). Also we used UBVRI photometry obtained at the 1.25m AZT-11 and wide-R photometry at the 2.6m Shain Telescope at the Crimean Astrophysical Observatory, Nauchny, Ukraine and unfiltered observations from the Arkansas Tech University Observatory. A journal of observations is presented in Table~\ref{tab1}.

\begin{table}
 \centering
 \caption{Journal of observations.}\label{tab1}
 \vspace*{1ex}
 \begin{tabular}{cccc}
  \hline
  UT date & HJD start & Length & Telescope \\
\hline
04.10.1997 & 50726.48273 & 0$^{\rm h}$44$^{\rm m}$ & ZTSh \\
02.11.1997 & 50755.57797 & 1$^{\rm h}$12$^{\rm m}$ & AZT11 \\
06.11.1997 & 50759.55697 & 1$^{\rm h}$45$^{\rm m}$ & AZT11 \\
01.02.1998 & 50846.30656 & 0$^{\rm h}$19$^{\rm m}$ & ZTSh \\
24.09.1998 & 51081.41335 & 2$^{\rm h}$46$^{\rm m}$ & AZT11 \\
16.12.1998 & 51164.34068 & 4$^{\rm h}$42$^{\rm m}$ & ZTSh \\
18.12.1998 & 51166.41945 & 4$^{\rm h}$00$^{\rm m}$ & ZTSh \\
19.12.1998 & 51167.40182 & 2$^{\rm h}$28$^{\rm m}$ & AZT11 \\
12.02.1999 & 51222.19297 & 3$^{\rm h}$30$^{\rm m}$ & AZT11 \\
17.03.1999 & 51255.25333 & 1$^{\rm h}$43$^{\rm m}$ & ZTSh \\
09.11.1999 & 51492.50222 & 1$^{\rm h}$29$^{\rm m}$ & ZTSh \\
10.11.1999 & 51493.52905 & 2$^{\rm h}$05$^{\rm m}$ & AZT11 \\
08.03.2000 & 51612.25129 & 2$^{\rm h}$02$^{\rm m}$ & ZTSh \\
01.10.2000 & 51819.42242 & 4$^{\rm h}$03$^{\rm m}$ & ZTSh \\
16.09.2001 & 52169.43919 & 3$^{\rm h}$07$^{\rm m}$ & ZTSh \\
18.09.2001 & 52171.46197 & 0$^{\rm h}$36$^{\rm m}$ & ZTSh \\
13.10.2001 & 52196.49104 & 2$^{\rm h}$01$^{\rm m}$ & ZTSh \\
16.10.2001 & 52199.44078 & 2$^{\rm h}$06$^{\rm m}$ & ZTSh \\
17.10.2001 & 52200.54955 & 1$^{\rm h}$33$^{\rm m}$ & ZTSh \\
03.10.2002 & 52551.59333 & 2$^{\rm h}$46$^{\rm m}$ & ZTSh \\
21.11.2003 & 52965.63002 & 1$^{\rm h}$47$^{\rm m}$ & ZTSh \\
15.09.2004 & 53264.62465 & 1$^{\rm h}$57$^{\rm m}$ & ZTSh \\
16.09.2004 & 53265.58709 & 2$^{\rm h}$40$^{\rm m}$ & ZTSh \\
01.10.2005 & 53645.59362 & 2$^{\rm h}$54$^{\rm m}$ & ZTSh \\
01.10.2005 & 53645.65030 & 1$^{\rm h}$43$^{\rm m}$ & ZTSh \\
30.11.2005 & 53705.40124 & 2$^{\rm h}$31$^{\rm m}$ & ZTSh \\
24.10.2006 & 54033.58360 & 2$^{\rm h}$04$^{\rm m}$ & ZTSh \\
14.02.2007 & 54146.34664 & 2$^{\rm h}$01$^{\rm m}$ & ZTSh \\
18.02.2007 & 54150.31628 & 4$^{\rm h}$03$^{\rm m}$ & ZTSh \\
10.01.2008 & 54476.34992 & 3$^{\rm h}$23$^{\rm m}$ & ZTSh \\
10.02.2008 & 54507.29841 & 3$^{\rm h}$13$^{\rm m}$ & ZTSh \\
08.03.2008 & 54534.23600 & 3$^{\rm h}$11$^{\rm m}$ & ZTSh \\
28.12.2008 & 54829.32144 & 0$^{\rm h}$10$^{\rm m}$ & ZTSh \\
16.01.2009 & 54848.55008 & 10$^{\rm h}$43$^{\rm m}$ & SCT 8-in \\
17.01.2009 & 54849.64778 & 6$^{\rm h}$34$^{\rm m}$ & SCT 8-in \\
20.01.2009 & 54852.69428 & 5$^{\rm h}$57$^{\rm m}$ & SCT 8-in \\
21.01.2009 & 54853.54531 & 7$^{\rm h}$24$^{\rm m}$ & SCT 8-in \\
22.01.2009 & 54854.54545 & 6$^{\rm h}$29$^{\rm m}$ & SCT 8-in \\
24.01.2009 & 54856.56511 & 4$^{\rm h}$44$^{\rm m}$ & SCT 8-in \\
02.02.2009 & 54865.65616 & 2$^{\rm h}$54$^{\rm m}$ & SCT 8-in \\
04.02.2009 & 54867.64652 & 7$^{\rm h}$40$^{\rm m}$ & SCT 8-in \\
16.02.2009 & 54879.56859 & 3$^{\rm h}$33$^{\rm m}$ & SCT 8-in \\
18.02.2009 & 54881.55606 & 7$^{\rm h}$27$^{\rm m}$ & SCT 8-in \\
19.02.2009 & 54882.60270 & 8$^{\rm h}$17$^{\rm m}$ & SCT 8-in \\
20.02.2009 & 54883.58367 & 7$^{\rm h}$02$^{\rm m}$ & SCT 8-in \\
21.02.2009 & 54884.59240 & 6$^{\rm h}$46$^{\rm m}$ & SCT 8-in \\
01.03.2009 & 54892.69511 & 4$^{\rm h}$34$^{\rm m}$ & SCT 8-in \\
19.08.2009 & 55062.53830 & 2$^{\rm h}$00$^{\rm m}$ & 50cm \\
29.08.2009 & 55072.55485 & 1$^{\rm h}$50$^{\rm m}$ & 50cm \\
24.09.2009 & 55098.52006 & 1$^{\rm h}$45$^{\rm m}$ & Meade \\
26.11.2009 & 55162.55397 & 11$^{\rm h}$05$^{\rm m}$ & SCT 10-in \\
27.11.2009 & 55163.61624 & 9$^{\rm h}$34$^{\rm m}$ & SCT 8-in \\
28.11.2009 & 55164.69199 & 3$^{\rm h}$11$^{\rm m}$ & SCT 8-in \\
30.11.2009 & 55166.68265 & 8$^{\rm h}$10$^{\rm m}$ & SCT 8-in \\
28.12.2009 & 55194.60363 & 9$^{\rm h}$33$^{\rm m}$ & SCT 10-in \\
\hline 
 \end{tabular}
\end{table}

\begin{table}
 \setcounter{table}{0}
 \caption{Journal of observations (continued).}\label{tab1c}
 \vspace*{1ex}
 \begin{tabular}{cccc}
  \hline
  UT date & HJD start & Length & Telescope \\
\hline
12.09.2010 & 55452.47810 & 3$^{\rm h}$47$^{\rm m}$ & VNT \\
23.09.2010 & 55463.41230 & 4$^{\rm h}$15$^{\rm m}$ & VNT \\
05.10.2010 & 55475.42049 & 5$^{\rm h}$31$^{\rm m}$ & Pupava \\
28.10.2010 & 55498.75629 & 5$^{\rm h}$20$^{\rm m}$ & SCT 16-in \\
29.10.2010 & 55499.71790 & 5$^{\rm h}$45$^{\rm m}$ & SCT 16-in \\
30.10.2010 & 55500.72724 & 5$^{\rm h}$39$^{\rm m}$ & SCT 16-in \\
31.10.2010 & 55501.72591 & 3$^{\rm h}$33$^{\rm m}$ & SCT 16-in \\
27.02.2011 & 55620.22137 & 5$^{\rm h}$22$^{\rm m}$ & VNT \\
26.08.2011 & 55800.45100 & 3$^{\rm h}$41$^{\rm m}$ & VNT \\
25.09.2011 & 55830.42984 & 5$^{\rm h}$11$^{\rm m}$ & VNT \\
26.09.2011 & 55831.47086 & 3$^{\rm h}$48$^{\rm m}$ & Pupava \\
27.09.2011 & 55832.35707 & 6$^{\rm h}$43$^{\rm m}$ & Pupava \\
29.09.2011 & 55834.49559 & 4$^{\rm h}$03$^{\rm m}$ & 50cm \\
30.09.2011 & 55835.42408 & 2$^{\rm h}$54$^{\rm m}$ & 50cm \\
02.10.2011 & 55837.47608 & 4$^{\rm h}$30$^{\rm m}$ & BAT \\
07.11.2011 & 55873.28985 & 9$^{\rm h}$36$^{\rm m}$ & VNT \\
08.11.2011 & 55874.36971 & 7$^{\rm h}$55$^{\rm m}$ & VNT \\
01.02.2012 & 55959.19697 & 6$^{\rm h}$46$^{\rm m}$ & VNT \\
12.02.2012 & 55970.23322 & 5$^{\rm h}$44$^{\rm m}$ & VNT \\
05.03.2012 & 55992.23876 & 4$^{\rm h}$18$^{\rm m}$ & VNT \\
16.11.2012 & 56248.26233 & 4$^{\rm h}$58$^{\rm m}$ & VNT \\
02.12.2012 & 56264.67947 & 7$^{\rm h}$53$^{\rm m}$ & SCT 8-in \\
04.12.2012 & 56266.63010 & 9$^{\rm h}$27$^{\rm m}$ & 5-inch TAK \\
01.01.2013 & 56294.35828 & 4$^{\rm h}$11$^{\rm m}$ & VNT \\
\hline 
 \end{tabular}
\end{table}
The photometric data were reduced using the C-Munipack software package. The final time series were obtained using the program MCV (I.L.Andronov, A.V.Baklanov \cite{mcv}) taking into account multiple comparison stars. Periodogram analysis was carried out using MCV and FDCN (Andronov \cite{fdcn}).

\section*{Determination of Maxima Timings}

For phase curves, we have used a preliminary value of the spin period of $P=0\fd0063131474$ and initial epoch $2449681.46389$ (Eq. (1) of Piirola et al. \cite{piirola08}). The phase curves for V, Rc observations are shown in Fig. 1 and 2. They exhibit a clear double--hump structure.
Using this value of the spin period, we have used a second-order trigonometric polynomial fit to the magnitudes 
\begin{eqnarray}
m(t)&=&C_1+C_2\cos(\omega t)+C_3\sin(\omega t)+\nonumber \\
&+&C_4\cos(2\omega t)+C_5\sin(2\omega t)
\end{eqnarray}
where $\omega=2\pi/P,$ and $P$ is period. The coefficients $C_3,$ $C_5$ describe differences between the two humps. However, the last two terms correspond to a "mean" shape of the pulse from two similar maxima. Thus, as the moment of maximum, we have used the moment of maximum $T_{02}$ of the first harmonic of the fit, i.e. 
\begin{eqnarray}
m_2(t)&=&C_4\cos(2\omega t)+C_5\sin(2\omega t)+\nonumber\\
&=&-R_2\cos(2\omega (t-T_{02}))
\end{eqnarray}

The minus sign before the value of semi-amplitude corresponds to a {\em maximum} of flux (i.e. the {\em minimum} of the brightness in stellar magnitudes). Obviously, in this method, the maximum of another pulse is shifted exactly by a value of $P/2.$ We preferred to use this parameter instead of individual maxima, which are determined with much worse accuracy.

\unitlength=1in
\begin{figure}[!h]
\centering
\includegraphics[width=83mm]{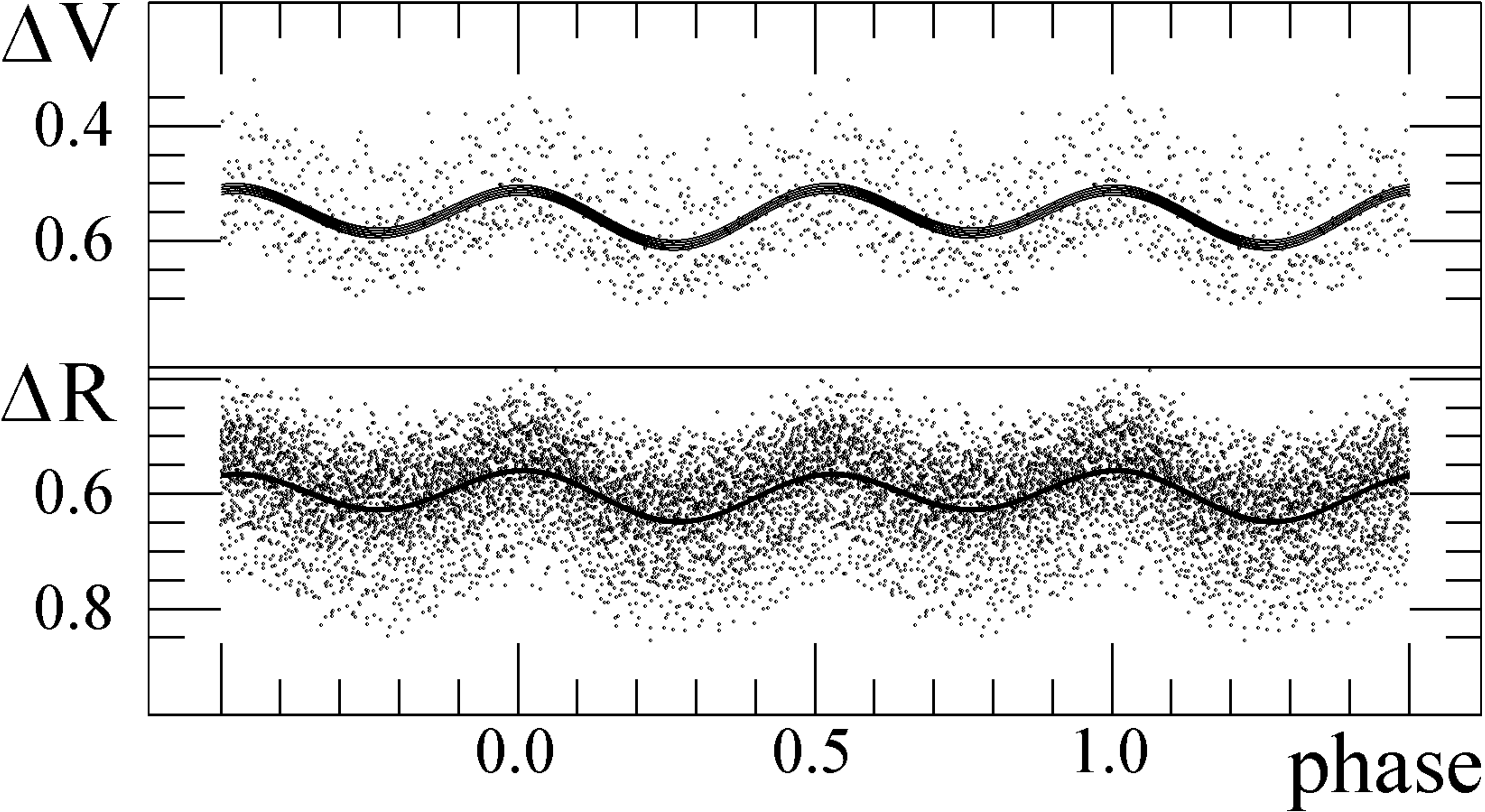}\\
\caption{Phase curve of the V405 Aur (circles) with a 2-nd order trigonometric polynomial approximation and $"\pm 1\sigma",$ $"\pm 2\sigma"$ error corridors.}\label{fig1}
\includegraphics[width=83mm]{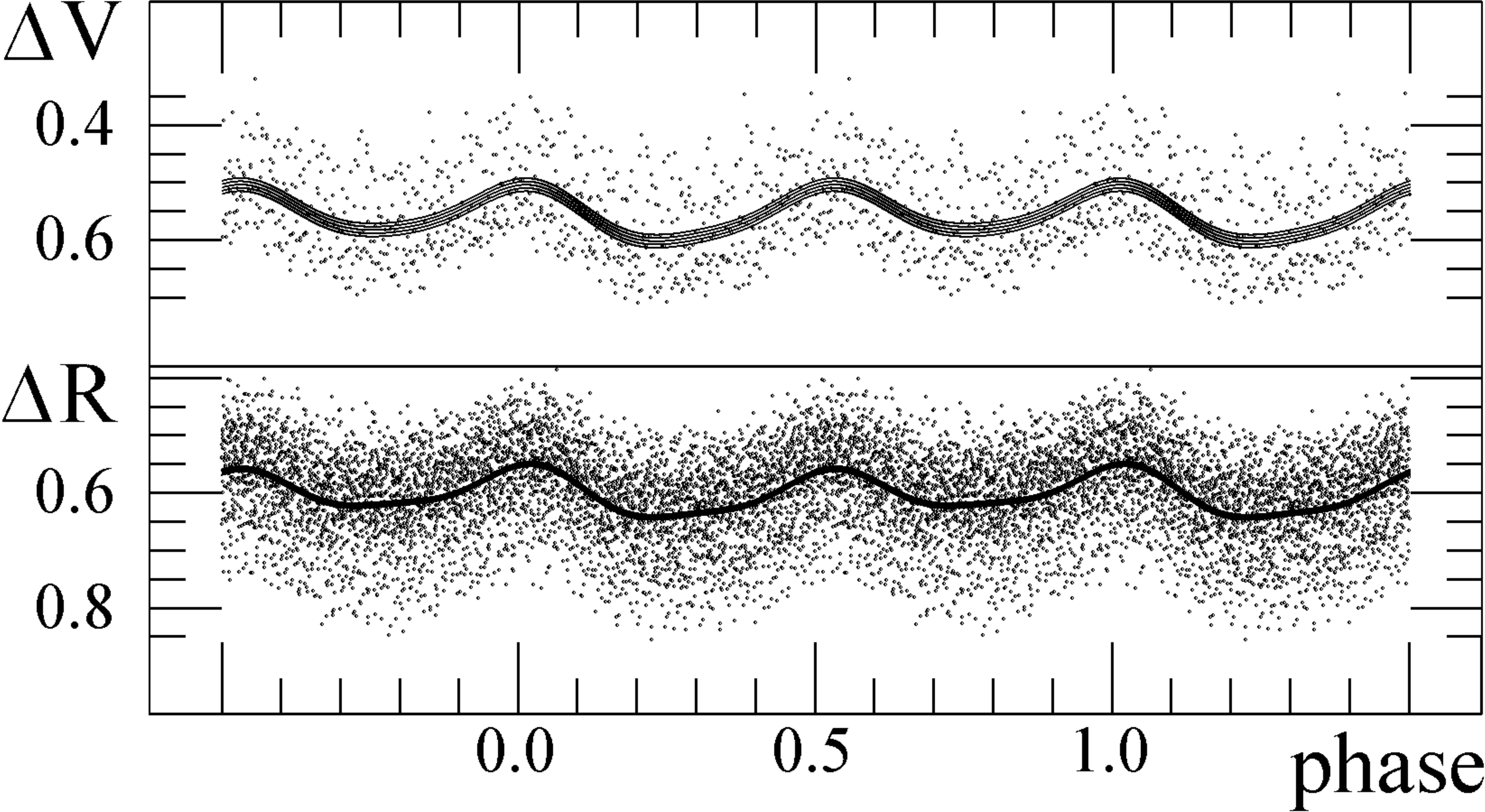}\\
\caption{Phase curve of the V405 Aur (circles) with a 4-th order trigonometric polynomial approximation and $"\pm 1\sigma",$ $"\pm 2\sigma"$ error corridors.}\label{fig2}
\end{figure}

Resulting "mean maxima timings" are listed in Table~\ref{tab2}. In an addition to our own time series, we have used the SuperWASP public archive \cite{wasp}. Initially, we have split these data into nightly runs. But, as the further analysis have shown, the scatter of phases is much larger than for our own data. Thus, for better statistics, we have merged the SuperWASP data into 6 intervals, and only timings with good accuracy were taken into account. Although, two moments (HJD 2455709.59139 and 2455909.37092) are also ``seasonal", corresponding not to a single night, but to {\em all} observational runs obtained in Kolonica in V and R, respectively.

\begin{table*}
 \begin{center}
 \caption{Spin maxima timings for V405 Aur}\label{tab2}
 \vspace*{1ex}
 \begin{tabular}{ccccccccccc}
\hline
HJD-2400000 & $\pm$ & Rem & ~ & HJD-2400000 & $\pm$ & Rem & ~ & HJD-2400000 & $\pm$ & Rem  \\
\hline
50726.49510 & 0.00003 & 7 & ~ & 53645.65783 & 0.00001 & 7 & ~ & 54854.67785 & 0.00005 & 8 \\
50755.60137 & 0.00013 & 6 & ~ & 53645.68671 & 0.00001 & 7 & ~ & 54856.66673 & 0.00008 & 8 \\
50755.60140 & 0.00008 & 6 & ~ & 53705.45918 & 0.00007 & 7 & ~ & 54865.71649 & 0.00012 & 8 \\
50755.60141 & 0.00006 & 6 & ~ & 54033.61713 & 0.00001 & 7 & ~ & 54867.80612 & 0.00008 & 8 \\
50755.60146 & 0.00011 & 6 & ~ & 54146.36630 & 0.00004 & 7 & ~ & 54879.64257 & 0.00009 & 8 \\
50755.60155 & 0.00010 & 6 & ~ & 54150.40657 & 0.00001 & 7 & ~ & 54881.70701 & 0.00011 & 8 \\
50759.59442 & 0.00014 & 6 & ~ & 54387.66949 & 0.00027 & 2 & ~ & 54882.77077 & 0.00007 & 8 \\
50759.59445 & 0.00014 & 6 & ~ & 54388.68261 & 0.00017 & 2 & ~ & 54883.73345 & 0.00008 & 8 \\
50759.59446 & 0.00008 & 6 & ~ & 54389.68600 & 0.00011 & 2 & ~ & 54884.73404 & 0.00005 & 8 \\
50759.59452 & 0.00007 & 6 & ~ & 54392.67580 & 0.00015 & 2 & ~ & 54892.78944 & 0.00006 & 8 \\
50759.59456 & 0.00006 & 6 & ~ & 54393.57216 & 0.00008 & 2 & ~ & 55162.78212 & 0.00003 & 8 \\
50846.31466 & 0.00003 & 7 & ~ & 54393.67628 & 0.00014 & 2 & ~ & 55163.82428 & 0.00006 & 8 \\
51081.47661 & 0.00004 & 6 & ~ & 54394.67363 & 0.00027 & 2 & ~ & 55164.75561 & 0.00013 & 8 \\
51081.47664 & 0.00011 & 6 & ~ & 54395.71194 & 0.00051 & 2 & ~ & 55166.85407 & 0.00005 & 8 \\
51081.47666 & 0.00005 & 6 & ~ & 54396.66273 & 0.00044 & 2 & ~ & 55194.80822 & 0.00004 & 8 \\
51081.47674 & 0.00008 & 6 & ~ & 54397.67291 & 0.00031 & 2 & ~ & 55452.55800 & 0.00005 & 3 \\
51081.47675 & 0.00007 & 6 & ~ & 54398.64800 & 0.00021 & 2 & ~ & 55452.55802 & 0.00005 & 3 \\
51164.44289 & 0.00001 & 7 & ~ & 54405.68025 & 0.00022 & 2 & ~ & 55463.50182 & 0.00007 & 3 \\
51166.50196 & 0.00001 & 7 & ~ & 54406.66561 & 0.00013 & 2 & ~ & 55463.50185 & 0.00006 & 3 \\
51167.45234 & 0.00005 & 6 & ~ & 54407.64707 & 0.00013 & 2 & ~ & 55475.53543 & 0.00009 & 3 \\
51167.45235 & 0.00003 & 6 & ~ & 54407.80797 & 0.00008 & 2 & ~ & 55498.86627 & 0.00005 & 8 \\
51167.45235 & 0.00005 & 6 & ~ & 54409.65134 & 0.00017 & 2 & ~ & 55499.89877 & 0.00004 & 8 \\
51167.45236 & 0.00003 & 6 & ~ & 54410.66438 & 0.00032 & 2 & ~ & 55500.84520 & 0.00003 & 8 \\
51167.45244 & 0.00005 & 6 & ~ & 54416.12864 & 0.00014 & 2 & ~ & 55501.79233 & 0.00005 & 8 \\
51222.26302 & 0.00006 & 6 & ~ & 54418.64771 & 0.00035 & 2 & ~ & 55620.33304 & 0.00002 & 3 \\
51222.26302 & 0.00007 & 6 & ~ & 54419.64116 & 0.00026 & 2 & ~ & 55709.59139 & 0.00004 & 3 \\
51222.26304 & 0.00009 & 6 & ~ & 54420.63655 & 0.00065 & 2 & ~ & 55800.52914 & 0.00007 & 3 \\
51222.26305 & 0.00006 & 6 & ~ & 54421.45335 & 0.00028 & 2 & ~ & 55830.53866 & 0.00004 & 3 \\
51222.26309 & 0.00007 & 6 & ~ & 54421.52664 & 0.00036 & 2 & ~ & 55831.54871 & 0.00007 & 3 \\
51255.30291 & 0.00003 & 7 & ~ & 54427.61780 & 0.00031 & 2 & ~ & 55832.49582 & 0.00005 & 3 \\
51492.53186 & 0.00011 & 7 & ~ & 54436.61510 & 0.00019 & 2 & ~ & 55834.57600 & 0.00009 & 4 \\
51493.57691 & 0.00005 & 6 & ~ & 54437.55549 & 0.00021 & 2 & ~ & 55834.57911 & 0.00007 & 4 \\
51493.57696 & 0.00006 & 6 & ~ & 54438.09861 & 0.00009 & 2 & ~ & 55835.48504 & 0.00010 & 4 \\
51493.57699 & 0.00004 & 6 & ~ & 54438.60984 & 0.00011 & 2 & ~ & 55835.48834 & 0.00012 & 4 \\
51493.57701 & 0.00004 & 6 & ~ & 54439.54781 & 0.00031 & 2 & ~ & 55837.58100 & 0.00008 & 5 \\
51520.27194 & 0.00020 & 1 & ~ & 54447.46696 & 0.00061 & 2 & ~ & 55837.58102 & 0.00013 & 5 \\
51612.29875 & 0.00001 & 7 & ~ & 54447.63452 & 0.00036 & 2 & ~ & 55873.49012 & 0.00002 & 3 \\
51819.50939 & 0.00001 & 7 & ~ & 54476.41950 & 0.00003 & 7 & ~ & 55874.53493 & 0.00003 & 3 \\
52169.50700 & 0.00001 & 7 & ~ & 54507.36966 & 0.00003 & 7 & ~ & 55909.37092 & 0.00002 & 3 \\
52171.47328 & 0.00003 & 7 & ~ & 54520.40288 & 0.00013 & 2 & ~ & 55959.33950 & 0.00003 & 3 \\
52196.53034 & 0.00002 & 7 & ~ & 54534.30477 & 0.00001 & 7 & ~ & 55970.35274 & 0.00004 & 3 \\
52199.45302 & 0.00004 & 7 & ~ & 54543.38921 & 0.00016 & 2 & ~ & 55992.32885 & 0.00003 & 3 \\
52200.57680 & 0.00002 & 7 & ~ & 54829.33040 & 0.00004 & 7 & ~ & 56248.36480 & 0.00004 & 3 \\
52551.65589 & 0.00002 & 7 & ~ & 54848.77188 & 0.00005 & 8 & ~ & 56264.83003 & 0.00006 & 8 \\
52965.67156 & 0.00002 & 7 & ~ & 54849.80093 & 0.00005 & 8 & ~ & 56266.81543 & 0.00005 & 8 \\
53264.66256 & 0.00001 & 7 & ~ & 54852.82184 & 0.00005 & 8 & ~ & 56294.44126 & 0.00004 & 3 \\
53265.64182 & 0.00001 & 7 & ~ & 54853.68357 & 0.00008 & 8 & ~ &  & & \\
 \hline 
 \end{tabular}\\
\end{center}
Remarks: (1) Data from NSVS. (2) Data from SuperWASP. (3) Observations at Kolonica Observatory.\\
	(4) VR photometry at 50cm. (5) VR photometry at the Beringer Automated Telescope (BAT).\\
	(6) UBVRI photometry at AZT-11. (7) WR (wide R) photometry at the 2.6m Shain Telescope.
	(8) Photometry by Jeff Robertson.\\
\end{table*}

\unitlength=1mm
\begin{figure}[!h]
\centering
\includegraphics[width=83mm]{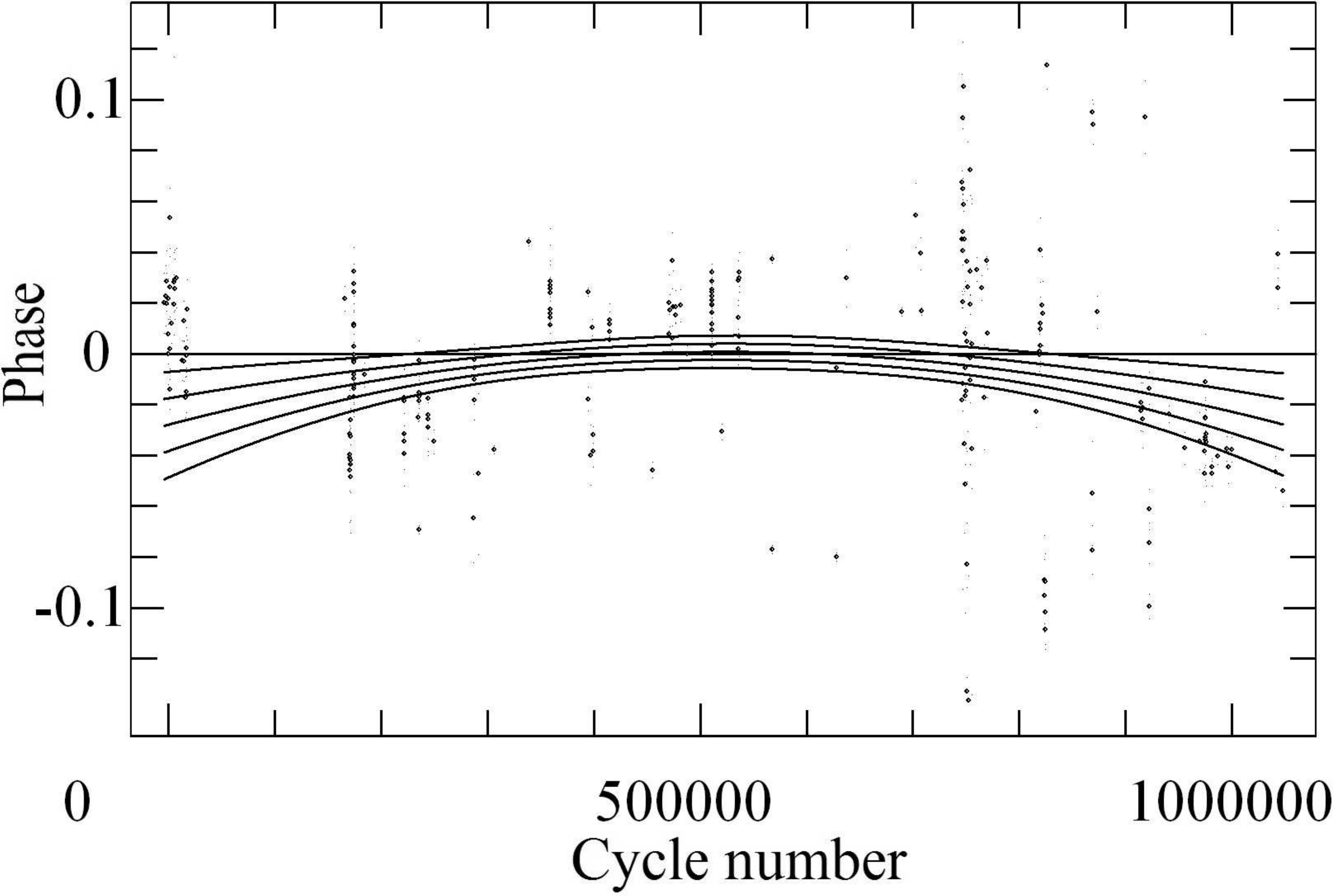}\\
\caption{Dependence of phases of maxima timings on cycle number of the spin period: circles - original observations, line - an approximation using 2-nd order polynomial fit with corresponding $\pm 1 \sigma$ and $\pm 2 \sigma$ error corridors.}\label{fig3}
\end{figure}

\unitlength=1mm
\begin{figure}[!h]
\centering
\includegraphics[width=83mm]{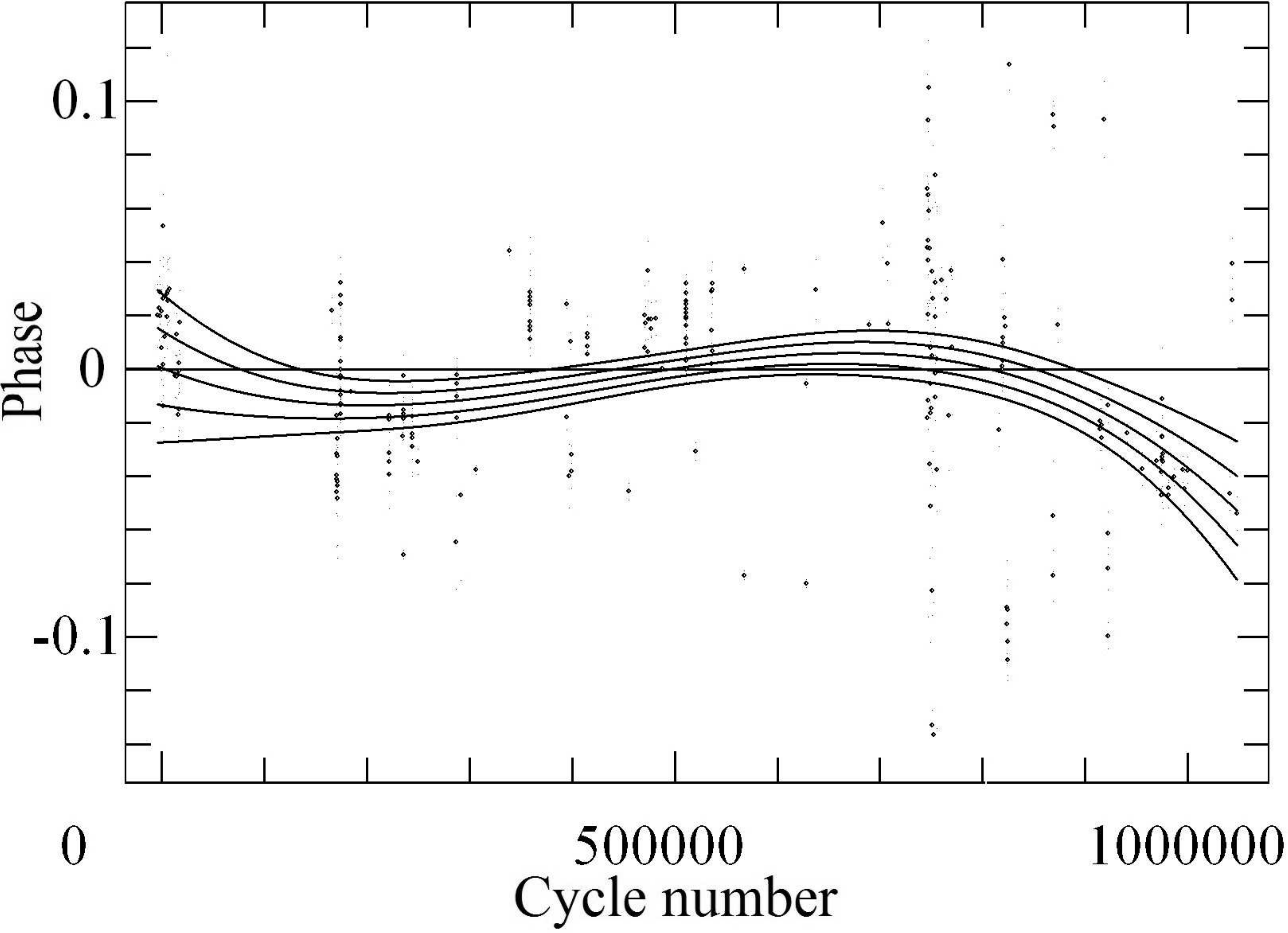}\\
\caption{Dependence of phases of maxima timings on cycle number of the spin period: circles - original observations, line - an approximation using 3-rd order polynomial fit with corresponding $\pm 1 \sigma$ and $\pm 2 \sigma$ error corridors.}\label{fig4}
\end{figure}

\unitlength=1mm
\begin{figure}[!h]
\centering
\includegraphics[width=83mm]{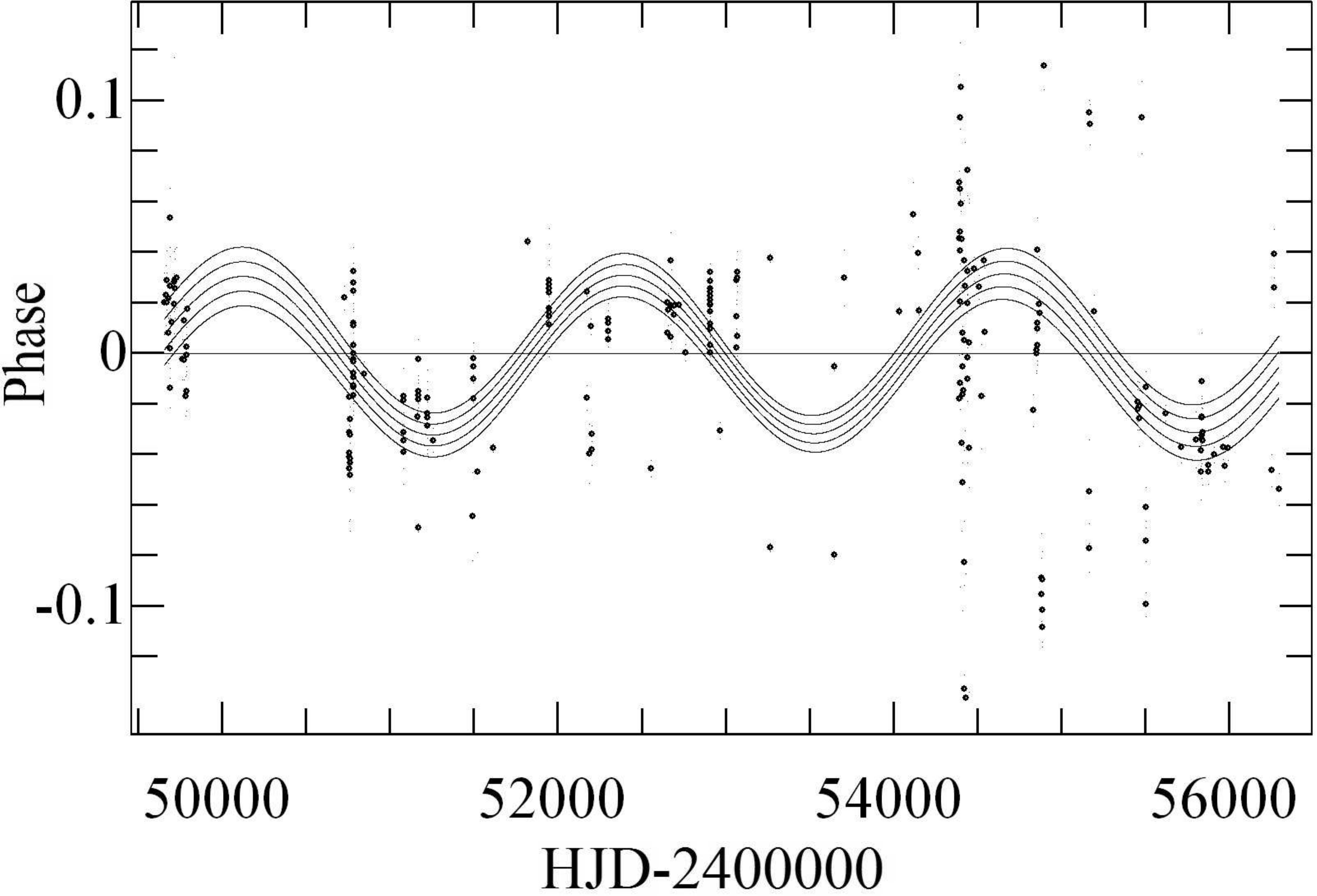}\\
\caption{Dependence of phases of maxima timings on HJD: circles - original observations, line - an approximation using 1-st order trigonometric and algebraic polynomial fits with corresponding $\pm 1 \sigma$ and $\pm 2 \sigma$ error corridors.}\label{fig5}
\end{figure}

It should be noted that unfortunately not all the authors published accuracy estimates of the timings. Thus we had to assign formal accuracy to their data. E.g. for the data published by Skillman \cite{skillman96} we assigned the value of $\sigma_\phi=0.0118,$ which was estimated as a standard error of phases from his list. For all other initial epochs published without error estimate we assigned a formal value of $\sigma_\phi=0.02$

\section*{"O-C" Analysis}

The O-C diagram for historical timings compiled by Piirola et al. \cite{piirola08}, maxima timings published in \cite{ashoka95}, \cite{skillman96}, \cite{evans04} and our own ones listed in the Table~\ref{tab2} is shown in Fig.~\ref{fig3}-\ref{fig5}. Contrary to a classical representation of the "O-C diagram" as a dependence of the timings from an ephemeris, i.e.
\begin{equation}
O-C=T-(T_0+P\cdot E)
\end{equation}
on the cycle number $E,$ we have used phases instead, i.e. $\phi=(O-C)/P.$ For a correct ephemeris, the phases should be concentrated near the zero value. In our case, the times of maxima estimated by a program correspond to a harmonic with a double frequency, thus it is expected that phases like $-0.5,$ $+0.5$ correspond to the same "zero", and thus one may make corrections for these values. For a complete set of "our$+$published" data, the phases seem to range from $-0.13$ to 0.11.

Contrary to a suggestion of Piirola et al. \cite{piirola08}, the points for the recent years show a distinct period {\it decrease}.

A simplest hypothesis is that the period has underwent change (approximately in 2007 $(E\approx 713491)$). 
We analysed separately photometric data, obtained at Vihorlat Astronomical Observatory in 2010-2013. As light curves in V and R filters had comparable amplitudes but different stellar magnitudes, so we subtracted the mean value of stellar magnitude from the data in each filter and joined it. We determined the new value of period and initial epoch that better corresponds all spin maxima timings in our observations.
\begin{equation}\label{form4}
T_{max}= HJD~2455882.470614(25) + 0.00631314602(46)\cdot E
\end{equation}

However, previous studies of intermediate polars argue for smooth period variations rather than period jumps, so we analysed other models for the period variations.

To decrease the error estimate of the initial epoch and the period, we computed the ephemeris for a different integer epoch $E_0,$ which is close to a sample mean of the observational values. In our case, $E_0=504600.$ A weighted fit to the phases of maxima $\phi$ (which are related to the traditional values of $(O-C)=\phi P$) leads to the following quadratic ephemeris:
\begin{eqnarray}\label{form2}
T_{max}&=& HJD~2452867.07807(2)\nonumber\\
&& + 0.006313147426(70)\cdot(E-E_{0})\\
&& - 659(233)\cdot 10^{-18}(E-E_{0})^{2}.\nonumber
\end{eqnarray}

The value of the quadratic term $Q$ reaches $2.8\sigma,$ so its deviation from zero is not statistically significant.
It doesn't fit our recent observations enough good (see Fig.~\ref{fig3}), so we calculated the 3-rd order weighted fit to the phases of maxima and got the following cubic ephemeris 

\begin{eqnarray}\label{form3}
T_{max}&=& HJD~2452867.07807(2)\nonumber\\
&& + 0.006313147760(131)\cdot(E-E_{0})\\
&& - 502(237)\cdot 10^{-18}(E-E_{0})^{2}\nonumber\\
&& - 239(80)\cdot 10^{-23}(E-E_{0})^{3}.\nonumber
\end{eqnarray}

It corresponds to all observations better (see Fig.~\ref{fig4}) and fits most recent observations showing a distinct negative trend.

Also we checked a hypothesis of periodic change of $O-C.$ We calculated the periodogram using the approximation combining a 1-st order trigonometric and a 1-st order algebraic polynomials (see Fig.~\ref{fig5}). 
The maximum peak at the periodogram corresponds to a period of $2268^{\rm d}=6.2$yr.
The corresponding fit is 
\begin{eqnarray}\label{form4}
\phi&=& - 0.00049(219) + 0.0000002(14) * (T-2452881)\nonumber\\
&& +0.0315(32) \cos(2\pi\cdot(T-2452389)/2268)
\end{eqnarray}

As these periodic variations are statistically significant (at a level of semi-amplitude of $9.7\sigma$), one may suggest a third body orbiting the inner binary system with a period of $\approx 6.2$yr, with a distance of the center of masses to the binary of 17.2 $\pm$ 1.8 light seconds, or $(5.15\pm0.53)\cdot 10^9$ meters). The corresponding mass function\cite{af} is $F(M)\approx0.09 M_\odot,$ so a third body may be a low-mass red dwarf.

\section*{Color Indices vs. Statistically Optimal Fit}

\begin{figure}[!h]
\centering
\includegraphics[width=83mm]{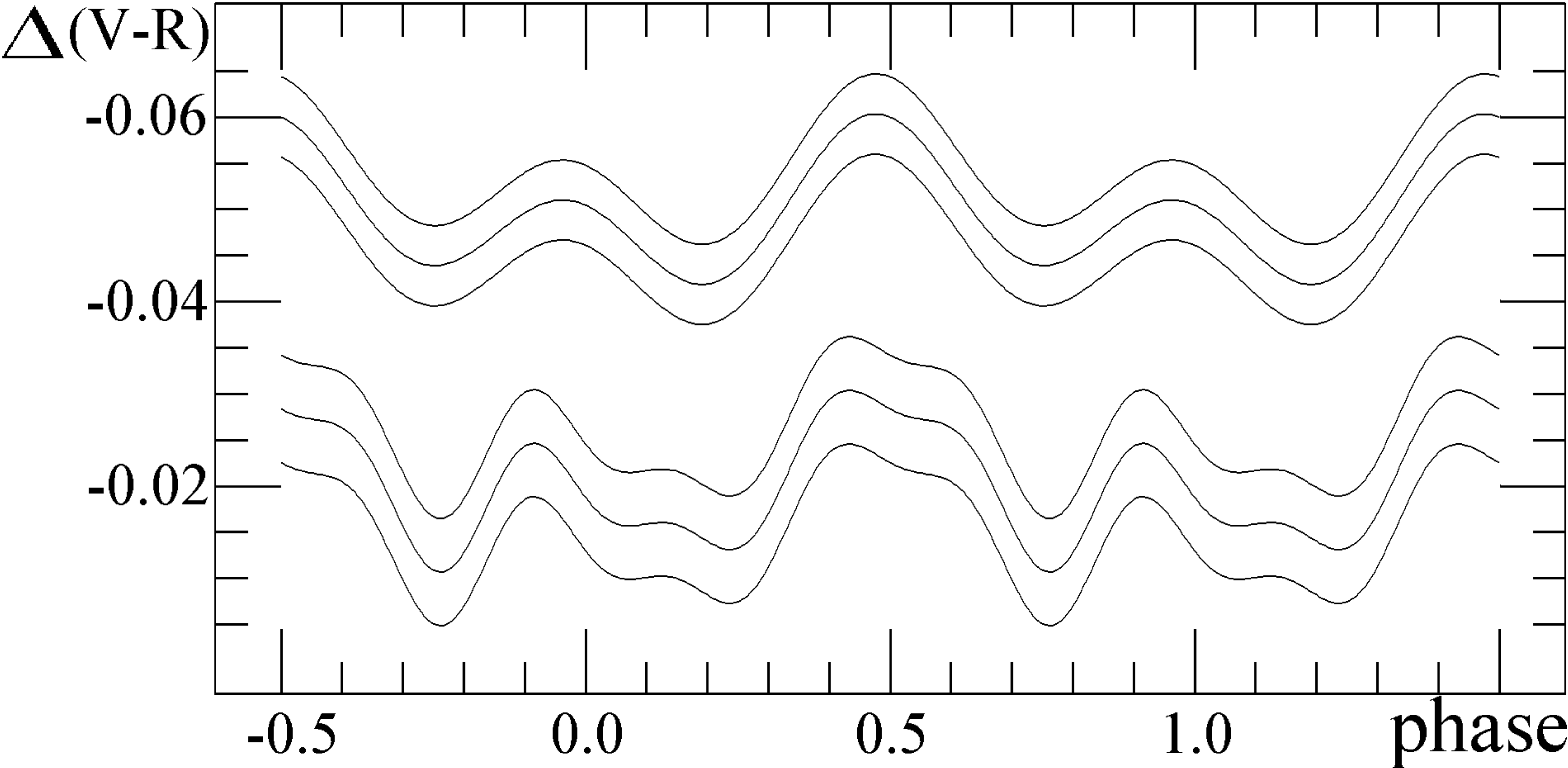}\\
\caption{The color index V-R phase curve for approximations using trigonometric polynomials of order 2 (upper, shifted by 0\fm03) and 4 (bottom) and corresponding $"\pm 1\sigma"$ corridor.}\label{fig6}
\end{figure}

Using the software FDCN \cite{fdcn} and the observations obtained in Kolonica, we determined that statistically optimal number of harmonics is 4. The phase 0 corresponds to a maximum preceding the primary (deeper) minimum, similarly to definition of other authors (e.g. \cite{allan95}). The best fit is shown in Fig.~\ref{fig2} for two filters V and R. To analyse variations of the color index in the instrumental system, for each phase we computed V-R and a corresponding error estimate 
\begin{equation}
\sigma_{V-R}=(\sigma_{V}^2+\sigma_{R}^2)^{1/2}
\end{equation}
The corresponding phase light curve is shown in Fig.~\ref{fig6}. It seems to be slightly asymmetric, with a maximum occurring slightly earlier than zero. The statistical error of the 4-th order trigonometric polynomial is much larger than for the upper curve, so the asymmetries in the shape are not statistically significant. For separate nights, the error estimates are even larger.
So finally we used a second-order trigonometric polynomial. Although the color indexes are nearly the same at phases of minima, the excess of the color index is twice larger at phase $\phi=0.5$ than at phase  $\phi=0.0.$ This is an interesting observational fact, which could be used for future modelling. A more detailed analysis of the curves of brightness and color index will be published elsewhere.

\unitlength=1mm

\section*{Discussion and Conclusions}

Period variations are frequently observed in intermediate polars and are typically detectable at a time scale of decades. Some objects do not show a statistically significant period change (e.g.1RXS $J180340.0+401214$ = RXJ 1803 =  V1323 Her  \cite{breus12}, some  
 show a period decrease 
 (e.g.  BG CMi \cite{kim04}, 
 EX Hya  \cite{breus12ex}, 
 \cite{andronov13ex}),
some show a period increase, but some show increase {\it and} decrease during a couple decades of observations (e.g. FO Aqr \cite{breus12fo}). 

From theoretical expectations, the spin periods of the white dwarf should be equal to some equilibrium value, which is equal to the period of ``Kepler" rotation of the inner accretion disk at a distance of the magnetospheric radius (\cite{warner}, \cite{hellier}). Period variations may be caused by changes of the accretion rate due to modulation of the mass transfer caused by magnetic activity of the red secondary (\cite{bianchini}, \cite{andronovshakun}) fluctuations of the orbital separation  \cite{andronovchinarova2001}, or precession of the magnetic white dwarf (which will be present either with constant, or variable accretion rate) \cite{andronov2005}.   For another object FS Aur in the same constellation, Tovmassian et al. (\cite{tovm03}, \cite{tovm07}) suggested precession of the magnetic white dwarf.

At time scales of decades, one may see only a part of the curve of cyclic variations. Thus apparently the "O-C" diagram may be not a ``wave", but a square (for smaller time intervals) or cubic parabola (for larger intervals).

An alternate model is a presence of a third body - a star or a massive planet. In this case, the theoretical ``O-C" diagram is a periodic wave (``mono--harmonic" for circular orbit or ``multi--harmonic" for elliptic orbit), which generally is superimposed onto a trend (cf. \cite{tsessevich1973}). Our analysis using the ``MCV" software \cite{mcv} shows that the harmonics are not statistically significant, so, assuming a circular orbit with a period of 6.2 yr, we obtained the value of the mass function of $f(M_3)=0.09M_\odot.$ As this is a lower limit of the mass of the third body, we may suggest that this is not a planet, but probably a red dwarf.

Contrary to the third-order polynomial, the sinusoidal fit shows a return to zero of phases at the end of the interval of observations. This difference will increase in future. Thus there may be a check-out, which model is correct, during even a year or two of subsequent observations.

{\it Acknowledgements.}
This paper is a part of the National project ``Ukrainian Virtual Observatory" \cite{ukrvo} and international project "Inter-Longitude Astronomy" \cite{ila}.

\end{document}